\documentclass{amsart}

\theoremstyle{definition}

\theoremstyle{remark}

\numberwithin{equation}{section}

\begin{document}

\title{New expressions for Laguerre and Hermite polynomials}

\author{H. Moya-Cessa}
\address{INAOE, Apdo. Postal 51 y 216, 72000, Puebla, Pue.,
Mexico}
\email{hmmc@inaoep.mx}

\subjclass[2000]{Primary }

\date{September 6th, 2008}



\begin{abstract}
New expressions for Laguerre and Hermite polynomials are shown.
They are based on operator algebras commonly used in quantum
mechanics.\end{abstract}

\maketitle

\section{Hermite polynomials}
Hermite functions are given by the expression \cite{A}
\begin{equation}
H_n(x)= (-1)^ne^{x^2}\frac{d^n}{dx^n}e^{-x^2} \label{herm}
\end{equation}
by doing $p=-id/dx$, we may rewrite them as
\begin{equation}
H_n(x)= (-i)^ne^{x^2}{p^n}e^{-x^2}
\end{equation}
given that $[x,p]f(x)=if(x)$ and considering the relation $e^{\mu
A}Be^{-\mu A}= B+\mu [A,B]+\mu^2/2![A,[A,B]]+ \dots$ we can cast
(\ref{herm}) as
\begin{equation}
H_n(x)= (-i)^n{(p+2ix)^n}. \label{hermite}
\end{equation}
We can add the Hermite polynomials given in (\ref{hermite}) to
obtain
\begin{equation}
\sum_{n=0}^{\infty}H_n(x)\frac{\alpha^n}{n!}=e^{-i\alpha(p+ix)}
\end{equation}
by applying the Baker-Hausdorff formula \cite{L},
$e^{A+B}=e^{-[A,B]/2}e^Ae^B$, valid when the commutators obey
$[A,[A,B]]=[B,[A,B]]=0$, we obtain
\begin{equation}
\sum_{n=0}^{\infty}H_n(x)\frac{\alpha^n}{n!}=e^{-\alpha^2+2\alpha
x}e^{-i\alpha p}=e^{-\alpha^2+2\alpha x}
\end{equation}
\subsection{Addition formula}
We want to apply the form obtained in (\ref{hermite}) to evaluate
the quantity
\begin{equation}
\nonumber H_n(x+y),
\end{equation}
we write it as
\begin{equation}
H_n(x+y)=(-i)^n{[-i\frac{d}{d(x+y)}+2i(x+y)]^n} \label{addin}
\end{equation}
by using the chain rule we have
\begin{equation}
\frac{d}{d(x+y)}=\frac{1}{2}\left(\frac{\partial}{\partial
x}+\frac{\partial}{\partial y}\right) \end{equation} so that we
may re-express (\ref{addin}) in the form
\begin{equation}
H_n(x+y)=
\left(\frac{-i}{\sqrt{2}}\right)^n(-i\frac{\partial}{\partial
\sqrt{2}x}+2i\sqrt{2}x-i\frac{\partial}{\partial \sqrt{2}y}
+2i\sqrt{2}y)^{n}
\end{equation}
by defining
\begin{equation}
p_X=-i\frac{\partial}{\partial X}, \qquad
p_Y-i\frac{\partial}{\partial Y}
\end{equation}
with $X=\sqrt{2}x$ and $Y=\sqrt{2}y$ we obtain
\begin{eqnarray}
H_n(x+y)=\frac{1}{2^{n/2}}\sum_{k=0}^{n}\left(
\begin{array} {l}
n\\
k
\end{array} \right)&&
(-i)^k{(p_X+2iX)^k}\\\nonumber &&\times(-i)^{n-k} (p_Y+2iY)^{n-k}
\end{eqnarray}
that by using (\ref{hermite}) adds to
\begin{eqnarray}
H_n(x+y)=\frac{1}{2^{n/2}}\sum_{k=0}^{n}\left(
\begin{array} {l}
n\\
k
\end{array} \right)H_k(\sqrt{2}x)H_{n-k}(\sqrt{2}y)
\end{eqnarray}

\section{Laguerre polynomials}
Laguerre polynomials are given by
\begin{equation}
L_n^{\alpha}(x)=
\frac{1}{n!}x^{-\alpha}e^x\frac{d^n}{dx^n}(e^{-x}x^{n+\alpha})
\end{equation}
that may be rewritten as
\begin{equation}
L_n^{\alpha}(x)=
\frac{1}{n!}x^{-\alpha}e^x(ip)^ne^{-x}x^{n+\alpha}
\end{equation}
by using that $e^xp^ne^{-x}=(p+i)^n$ we can cast them into
\begin{equation}
L_n^{\alpha}(x)=
\frac{1}{n!}x^{-\alpha}\left(\frac{d}{dx}-1\right)^nx^{n+\alpha}
\label{Lag}
\end{equation}
or
\begin{equation}
L_n^{\alpha}(x)= \frac{1}{n!}x^{-\alpha}\sum_{m=0}^n \left(
\begin{array} {l}
n\\
m
\end{array} \right)(-1)^{n-m}\frac{d^m}{dx^m}x^{n+\alpha}
\end{equation}
because
\begin{equation}
\frac{d^m}{dx^m}x^{n+\alpha}=\frac{(n+\alpha)!}{(n+\alpha-m)!}x^{n+\alpha-m}
\end{equation}
so we obtain the usual form for Laguerre polynomials:
\begin{equation}
L_n^{\alpha}(x)= \sum_{k=0}^n \left(
\begin{array} {l}
n+\alpha\\
n-k
\end{array} \right)(-1)^k\frac{x^k}{k!}.
\end{equation}
\section{Conclusions}
We have given new forms for Hermite polynomials (\ref{hermite})
and Laguerre polynomials (\ref{Lag}) that may be used to add these
polynomials in an easier form.

\end{document}